# Advanced Optics with Laser Pointer and Metersticks

**Roman Ya. Kezerashvili**

*New York City College of Technology, The City University of New York*
*300 Jay Street, Brooklyn NY, 11201*
*Email: Rkezerashvili@citytech.cuny.edu*

**Abstract**

We are using a laser pointer as a light source, and metersticks as an optical branch and the screen for wave optics experiments. It is shown the setup for measurements of wavelength of laser light and rating radial spacing of the CD, diffraction on a wire and a slit, observation of a polarization of light and observation of a hologram.

## 1. Introduction

Laser pointers also know as laser penlights, have become very affordable recently due to new developments in laser technology. They are widely available at electronic stores, novelty shops, through mail order catalogs and by numerous other sources. They are in the price range from $1 to $30 as other electronic toys and are being treated as such by many parents and children. Pointers are used for other purposes such as the aligning of other lasers, laying pipes in construction, and as aiming devices for firearms.

Laser pointer can be use to observe the interference, diffraction and polarization of light in college physics laboratory. Laser pointers can be used to produce holograms[1-2]. It been opinions it couldn't be done because of the short coherence of the beam and that laser pointer's beam was not polarized. But it was practically proved that a laser pointer could be used not only to observe a hologram but to produce a high quality transmission and reflection display hologram[1-3].

We suggested a series of classic experiments with a laser pointer as a light source, and using metersticks as an optical branch and the screen to measure the interference and diffraction patterns. The detail procedures of suggested experiments are given in Ref. 4.

## 2. Determination of wavelength of laser Light

A wave phenomenon, which occurs when two or more waves overlap in the same region of space at the same time and form an interference pattern, is known as the interference of waves. When two waves of the same frequency but of different phases combine, the resultant wave is a wave of the same frequency, the amplitude of which depends on the phase difference. If the phase difference is 0 or an integer times $2\pi$, the waves are in phase and interfere constructively. For the constructive interference, the resultant amplitude equals the sum of the individual amplitudes, and the intensity has a maximum. If the phase difference is $\pi$ or any odd integer times $\pi$, the waves are out of phase and interfere destructively. For destructive interference the resultant amplitude is the difference between the individual amplitudes, and the intensity has a minimum. If the two sources have the same amplitude, the destructive interference results in the zero net amplitude. The phase difference between two waves is often the result of difference in path length traveled by the two waves. If waves from two sources with the same wavelength arrive at the same point together with exactly the same phase, then the condition for maximum constructive interference is that the path length of two waves must be identical or else differ by an integer multiple of the wavelength, that is

$$D_1 - D_2 = m\lambda, \qquad m = 0, 1, 2, 3, ... \text{ (constructive interference)}, \qquad (1)$$

where $D_1$ and $D_2$ are the path lengths of the waves from their source to the point $P$ (Fig. 1). If the distances $D_1$ and $D_2$ are quite large compared to the separation $d$ between the sources, we can write the conditions for the constructive and destructive interferences in terms of the angle $\theta$ and the distance of separation $d$. When the distance $d$ between the light sources and the plane containing the observation point is much greater than $L$, two paths $D_1$ and $D_2$ are nearly parallel and the path difference is approximately $d\sin\theta$, as shown in Fig.1. This result gives equation for determining the constructive interference of the resultant wave at point $P$:

$$d\sin\theta_m = m\lambda, \qquad m = 0, 1, 2, 3, ... \text{(constructive interference)}, \qquad (2)$$

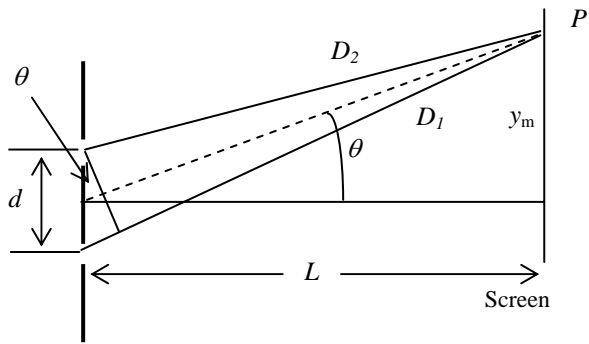

**Fig.1.** Two slits act as coherent sources of light for the observation of interference. Because the screen is very far away compared with the slit separation $d$, the rays from the slits to the point $P$ on the screen are considered approximately parallel and the path difference between the two rays is $d \sin \theta$.

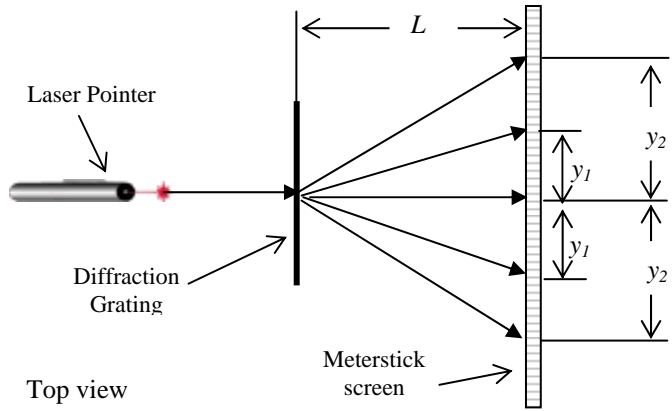

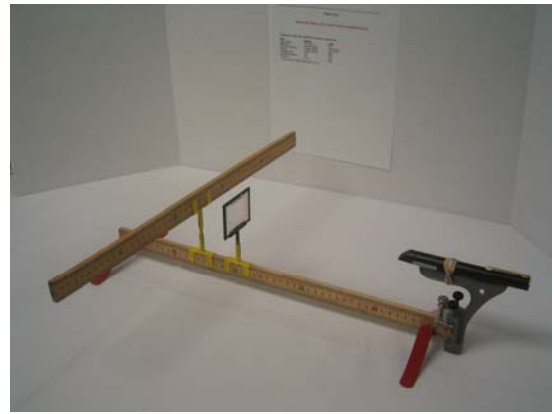

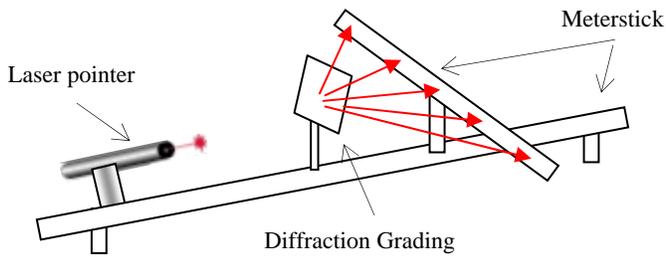

**Fig.2.** Interference. Setup for measurement of the wavelength of a laser light.

As it is shown in Fig. 1, the distance $y_m$ measured along the screen from the central bright point to the $m$th bright fringe is related to the angle $\theta$ by

$$\sin\theta_m = \frac{y_m}{\sqrt{L^2 + y_m^2}} \qquad (3)$$

Substituting this into equation (2) we obtain

$$d\frac{y_m}{\sqrt{L^2 + y_m^2}} = m\lambda. \qquad (4)$$

Solving equation (4) for wavelength we get

$$\lambda = \frac{d}{m}\frac{y_m}{\sqrt{L^2 + y_m^2}}, \qquad m = 1, 2, 3,... \text{ (constructive interference)}. \qquad (5)$$

Thus, for the known distance of separation $d$, by measuring the distance $y_m$ along the screen from the central bright point to each of the $m$th bright fringe and the distance $L$ from the slit to the screen, we can determine the wavelength of the monochromatic light. We determine the wavelength of the laser light using diffraction grating. In the first part of this experimental activity we will be using a monochromatic laser beam, which is incident normally on a transmission diffraction grating. By observing the interference pattern and measuring the distances $y_m$ between the central bright maximum and each of the $m$th bright fringe and the distance $L$ from the diffraction grating to the screen, we will determine the wavelength of the laser light from equation (5).

### 3. Determination of the grating radial spacing of the CD

By solving equation (5) for the distance of separation $d$ we obtain

$$d = m\lambda \frac{\sqrt{L^2 + y_m^2}}{y_m}, \qquad m = 1, 2, 3,... \text{ (constructive interference)}. \qquad (6)$$

Equation (6) allows us to determine the distance of separation between the light sources. That means we can find a grating spacing by using a light source of a known wavelength and measuring the distance $y_m$ along the screen from the central bright point to each of the $m$th bright fringe and the distance $L$ from the slit to the screen.

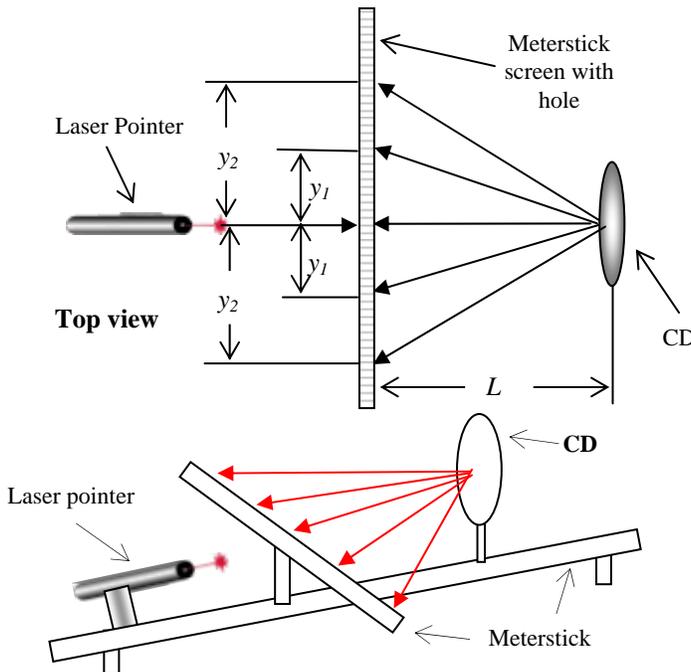
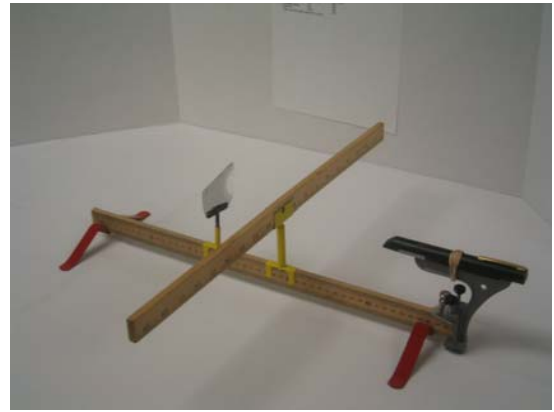

**Fig. 2.** Interference. Setup for measurement a grating space of a CD.

The rainbow-colored reflections that you can see from the surface of a CD are the reflection grating effects. The grooves are tiny pits 0.1 μm deep on the surface of the disc, with the uniform radial spacing of $d=1.6$ μm. The reflection grating aspect of the CD is merely an aesthetic side benefit. The interference pattern produced on the screen on a large distance from the grating is due to a large number of equally spaced light sources. The interference maxima are at the angle $\theta$ given by equation (3). In this part of the experiment the monochromatic laser beam will incident normally on a reflection diffraction grating. We will be using a CD as the reflection diffraction grating. By measuring the distances $y_m$ between the central bright maximum and each of the $m$th bright fringe, the distance $L$ from the CD to the screen and using the wavelength of the laser light found previously, we will determine the grating radial spacing of the CD from equation (6).

## 4. Diffraction on a wire

In this experiment a narrow slit or a thin wire is inserted in the path of a laser pointer beam and diffracted spots are projected on a meterstick-screen as it is shown in Fig. 4. The laser beam is assumed to be very parallel rays of light, which means that the waves are plane light waves, passes though a slit. Plane waves are diffracted by the slit and light rays fall on the screen, which is very far away, so the rays heading for any point are very nearly parallel before they meet at the screen. This approximates the conditions for Fraunhofer diffraction. At the screen a central bright region is formed and a series of symmetrical dark and bright fringes can be observed. A general condition for a dark fridge of the diffraction pattern is

$$a \sin \theta_m = m\lambda, \quad m = \pm 1, \pm 2, \pm 3,..., \quad \text{(Dark Fringe)} \quad (7)$$

where $a$ is the width of a slit and $\theta_m$ is the angle between the position of the dark fringe of $m$ order and central bright fringe. The central bright fringe occurring at $\theta=0$ and is the brightest, secondary and so on bright fringes occurring on both side of the central maximum. However, the intensity of those maxima diminish more rapidly. From the Fig. 4 we see that for the small angle $\theta_m$ we can replace $\sin \theta_m$ by the expression

$$\sin \theta_m \approx \frac{d_m}{2L}, \quad m = 1, 2, 3,..., \quad (8)$$

where $d_m$ is the distance between two symmetrical dark fringes of the same order $m$, and $L$ is a distance between the wire and the screen. Substitution of equation (8) into equation (7) gives the width of the slit as

$$a = \frac{2m\lambda L}{d_m}, \qquad m = 1, 2, 3, 4, ..., \qquad (9)$$

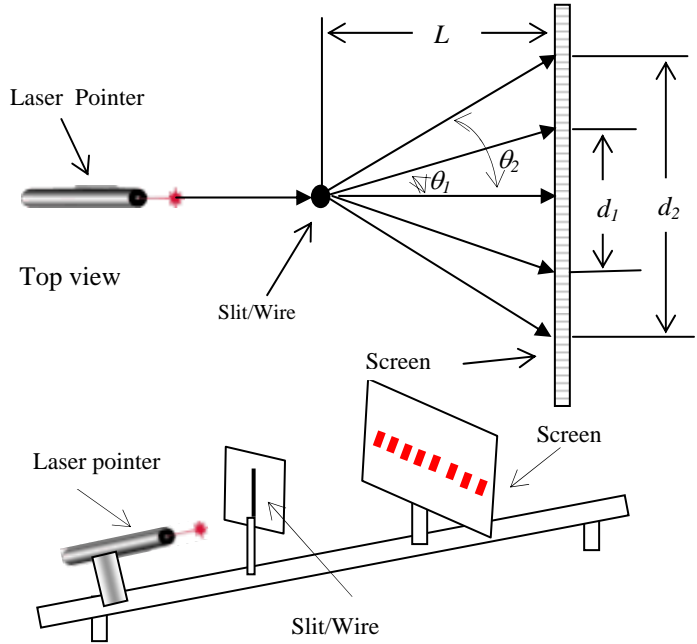
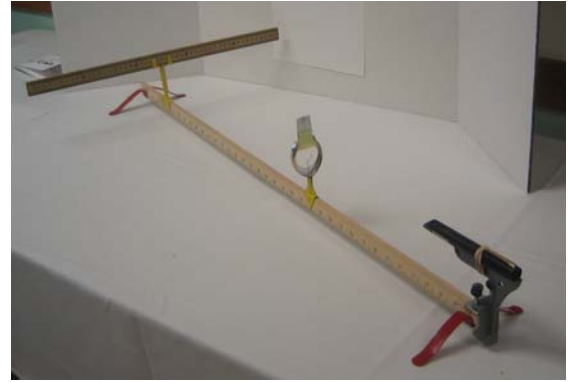

**Fig. 4.** Diffraction from a slit or on a wire

Therefore, we can find the width of very narrow slit by measuring the distances $d_m$ between two symmetrical dark fringes of the same order $m$, and the distance $L$.

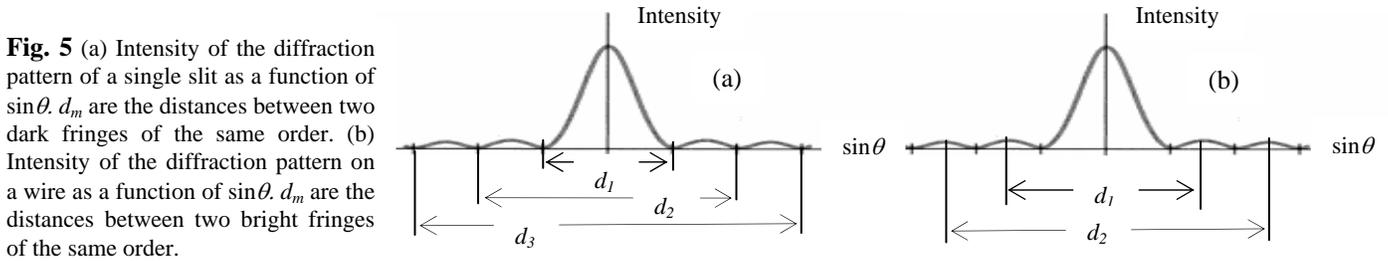

**Fig. 5** (a) Intensity of the diffraction pattern of a single slit as a function of $\sin\theta$. $d_m$ are the distances between two dark fringes of the same order. (b) Intensity of the diffraction pattern on a wire as a function of $\sin\theta$. $d_m$ are the distances between two bright fringes of the same order.

Now let us consider the diffraction on an obstacle. Replace the narrow slit slide by a wire holding slide. You can observe the diffraction pattern. The central bright fringe occurring at $\theta=0$ and is the brightest, secondary and so on bright fringes occurring on both side of the central maximum. A general condition for a bright fridge of the diffraction pattern is

$$a\sin\theta_m = m\lambda, \quad m = \pm 1, \pm 2, \pm 3,..., \quad \text{(Bright Fringe)} \qquad (10)$$

where $a$ is the diameter of a wire and $\theta_m$ is the angle between the position of the bright fringe of $m$ order and central bright fringe. We can determine the diameter of the wire using equation (10) and as a result it will be yield to equation (9) only now distance $d_m$ is the distance between two symmetrical bright fringes of the same order $m$. Thus, we can find the diameter of the thin wire by measuring the distances $d_m$ between two symmetrical bright fringes of the same order $m$, and the distance $L$. In this experiment using a laser pointer you can observe the sharp diffraction pattern for the wire of the gages from 34 to 48 (0.16 -0.03 *mm*).

## 5. Observation of a Polarization of Light

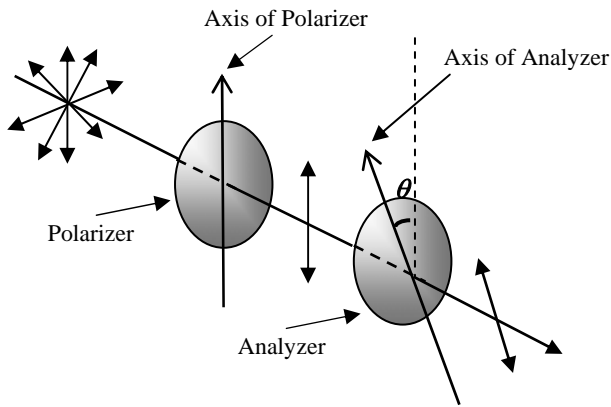

**Fig. 6.** Unpolarized light becomes polarized after passing through the first Polaroid (polarizer). The plane of polarization is changed and the intensity is reduced by factor $\cos^2\theta$ as the light passes the second Polaroid (analyzer) oriented at angle $\theta$ relative to the first.

If an unpolarized light is incident on a polarizing material, the transmitted light is linearly polarized in the direction parallel to the transmission axis of the polarizer, as in Fig.6. The polarizer transmits half of an intensity of the incident light, while the other half is absorbed by the polarizer. This fact is known as the *one-half rule*. We can use this rule only when the light reaching the polarizer is unpolarized. If the light is linearly polarized when it is incident on the second polarizer, as in Fig.6, then the second polarizer examines or analyzes the incoming light and can tell us something about the nature of this polarized light. The second polarizing material thus called an analyzer. In other words, when two polarizing materials are placed in succession in a beam of light as on Fig.6, the first is called the polarizer and the second - the analyzer. The amount of the light transmitted by the analyzer depends on the angle $\theta$ between its transmission axis and the direction of the axis of the polarizer. The observations show that the amplitude of the electric field of the light is reduced by the second polarizer (analyzer) by factor $\cos\theta$. Since the intensity of the light, which is the energy transmitted per unit time and area, is proportional to the amplitude squared, the intensity of the light transmitted by both polarizer and analyzer will be given by

$$I = I_0 \cos^2 \theta . \qquad (11)$$

In equation (11) $I_0$ is the intensity of the light incident on the analyzer ($I_0$ is the maximum intensity of the light transmitted at $\theta=0$), and $I$ is the intensity of the light transmitted by both polarizer and analyzer. If the transmission axis of the polarizer and analyzer are perpendicular to each other, no light gets through. Equation (11) is known as Malus's law. This law applies to any two polarizing elements, transmission axes of which make an angle $\theta$ with each other.

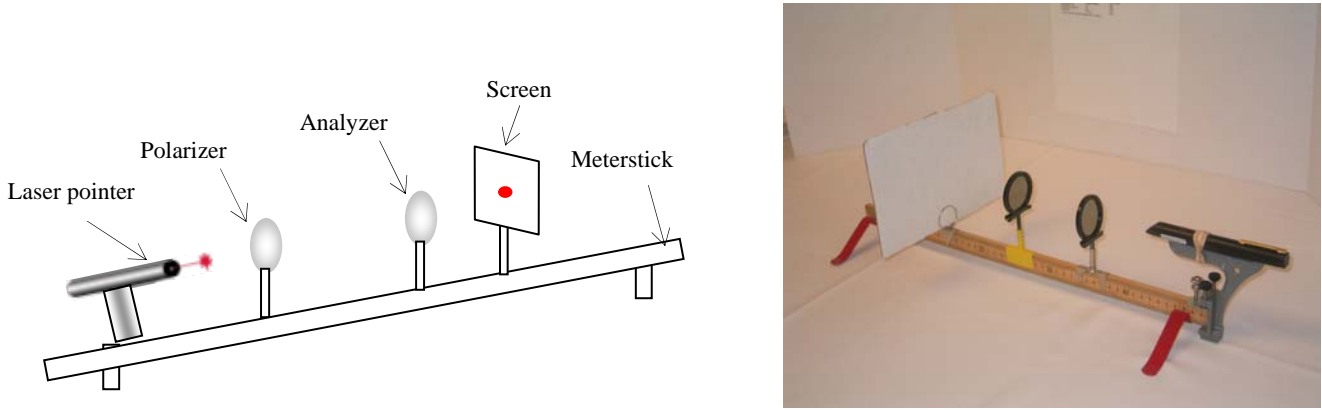

**Fig. 7.** Setup for observation of the polarization of light

### 6. Observation of a Hologram

One of the most familiar applications of laser is a process for producing three-dimensional images.

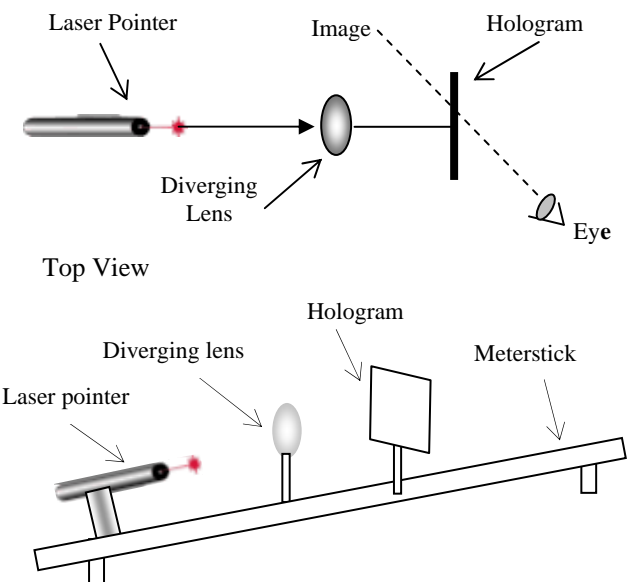

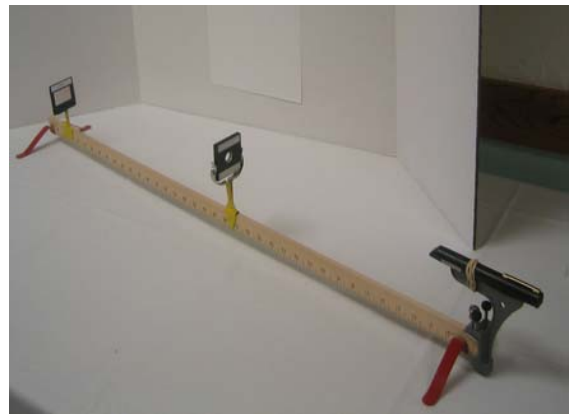

**Fig. 8.** Setup for observation of the hologram

To produce a holographic image, the laser beam from the laser pointer passes through the center of the diverging lens (from -6*mm* to -25*mm* focal length, double concave lens) and as a result spreads the beam to uniformly illuminate the transparent hologram. The basic setup for a single beam transmission hologram is shown in Fig. 8. The distance between the diverging lens and the hologram holder is approximately 1 *m* (the distance depends on the focal length of the diverging lens) to get uniform illumination.